# The self-oscillation paradox in the flight motor of *D. melanogaster*

Arion Pons[*]

Division of Fluid Dynamics, Department of Mechanics and Maritime Sciences,
Chalmers University of Technology, Gothenburg, Sweden

**Abstract:** Tiny flying insects, such as *Drosophila melanogaster*, fly by flapping their wings at frequencies faster than their brains are able to process. To do so, they rely on self-oscillation: dynamic instability, leading to emergent oscillation, arising from muscle stretch-activation. Many questions concerning this vital natural instability remain open. Does flight motor self-oscillation necessarily lead to resonance – a state optimal in efficiency and/or performance? If so, what state? And is self-oscillation even guaranteed in a motor driven by stretch-activated muscle, or are there limiting conditions? In this work, we use data-driven models of wingbeat and muscle behaviour to answer these questions. Developing and leveraging novel analysis techniques, including symbolic computation, we establish a fundamental condition for motor self-oscillation common to a wide range of motor models. Remarkably, *D. melanogaster* flight apparently defies this condition: a paradox of motor operation. We explore potential resolutions to this paradox, and, within its confines, establish that the *D. melanogaster* flight motor is likely not resonant with respect to exoskeletal elasticity: instead, the muscular elasticity plays a dominant role. Contrary to common supposition, the stiffness of stretch-activated muscle is an obstacle to, rather than an enabler of, the operation of the *D. melanogaster* flight motor.

## 1. Introduction

The flight motors of Dipterans, such as *Drosophila* spp., are fascinating natural examples of self-oscillation. These motors are complex set of muscles, neural control systems, and exoskeletal structures – including wings, halteres, and transmission mechanisms – that together generate controlled and powered flight. The primary flight muscles within these motors are remarkable in that they are asynchronous: they are subjected to neural activation as infrequently as every 40th wingbeat [1]. In the intervening periods, the muscles continue to oscillate, and the wings continue to beat, through the phenomenon of stretch-activation: the non-neural delayed activation of asynchronous muscles in response to changing strain [1–3]. The Dipteran flight motor, under stretch activation, is a system that converts a form of energy that has no inherent periodicity – chemical energy, in adenosine triphosphate, ATP – into periodic oscillation. As such, it is a dynamically-unstable, self-oscillating system [4,5].

Early on during the study of stretch-activation in insect flight motors, connections were made between motor self-oscillation and motor resonance. The landmark experiments of Machin and Pringle [4], on asynchronous muscles attached to a lightly-damped load system, demonstrated characteristics of the self-oscillatory response that would suggest alignment with the system natural and/or resonant frequency [6]. Subsequently, self-oscillation and resonance have become blended terms in the literature [7–9], and the Dipteran flight motor is widely depicted as resonant [10–12]. However, a remarkable feature of the results Machin and Pringle [4] is now seldom noted: in a general system, there is no particular reason why self-oscillation and resonance should align [13]. Indeed, both older [14] and recent [5] analyses of insect self-oscillation models indicate that convergence to a resonant frequency is not a guaranteed property of the motor system. The identification of multiple resonant frequencies within even simple models of the Dipteran flight motor – frequencies split by damping and by the distribution of stiffness/elasticity [6] – adds further colour to this picture. Is self-oscillation still resonant under *in vivo* heavy damping conditions? [12] If so, resonant in what sense? [6] And what, indeed, is the quantitative mechanism by which motor self-oscillation arises from stretch-activation? [15]

This work seeks to answer these questions. Using literature-reported *ex vivo* data from *D. melanogaster* asynchronous flight muscles, and data-driven models of the *D. melanogaster* wingbeat and flight motor, and several novel analysis techniques, we characterise the self-oscillation behaviour of this Dipteran's flight motor. In §§2-3 we develop data-driven models of the wingbeat and muscular behaviour, respectively, utilising a new method for characterising muscular impulse-response data. In §4 we analyse these motor models using symbolic computation, and encounter a fundamental and pervasive paradox: self-oscillation at observed wingbeat amplitudes should not be possible. This paradox holds irrespective of the motor model; motor parameters, including exoskeletal stiffness distribution, exoskeletal damping, and quantity of muscle; and the muscular data source. Reported muscular *ex vivo* behaviour – from both sinusoidal and strain rate impulse response testing – is simply too stiff to match wingbeat requirements. As such, these results challenge the assertion that the high stiffness of asynchronous flight muscles is a feature enabling motor self-oscillation [10,16,17]: in contrast, this stiffness is an obstacle to self-oscillation in heavily-damped insects such as *D. melanogaster*. They also qualify the relationship between self-oscillation and resonance. In §5 we explore this relationship, and both confirm and extend the conclusions of Machin and Pringle [4]. For lightly-damped models of the Dipteran flight motor, self-oscillation is resonant with respect to the thoracic exoskeleton. However, for *in-vivo* levels of motor damping, resonant behaviour is likely driven by the elasticity of the muscles. Altogether, as we discuss finally in §6, our results provide new insight into the dynamical behaviour of flight muscles: the connection, and distinction, between self-oscillation and resonance in fully integrated models of the Dipteran flight motor and musculature. And the new analysis techniques that we develop and apply, including symbolic computation routines, and methods for the analysis of impulse response data, open up avenues for wider meta-analyses of flight motors across multiple species. The prospects for further integrated studies – and for a final answer to the question of whether Dipteran flight motors really are resonant – are positive.

## 2. Modelling flight motor oscillation

### 2.1. Modelling flight motor dissipation

The study of flight motor self-oscillation is predicated on a model of the motor as an oscillator. In constructing such a model several assumptions are typically made (Fig. 1). *First*: the single-degree-of-freedom (1DOF) assumption – the treatment of a 1DOF drivetrain pathway between a reference muscular strain and a wingbeat kinematic variable [12,15]. The presence of series elasticity – between the muscles and each wing – may technically extend the model to 2DOF [6,15], but 1DOF analyses remain possible. 1DOF models represent reductions of the real continuum mechanics of the exoskeletal structure [18]. Making the 1DOF assumption typically involves the assumption of a rigid wing [12], though these assumptions are properly distinct. There is no preclusion of the use of wing deformation for aerodynamic modelling in a 1DOF model, *cf.* [19,20]; and the elastic effect of wing deformation could be included to first order via series elasticity [6,15]. *Second*: the planar assumption – the analysis of the wingbeat stroke angle as the primary DOF for muscular forcing [12,15]. The planar assumption also does not preclude the use of non-planar wingbeat kinematics for aerodynamic modelling. It posits only that out-of-plane kinematics do not impact the forcing requirements of the primary flight muscles – either, as comparatively insignificant components of muscular power consumption [22,23], or, as generated by passive non-muscular effects [24,25].

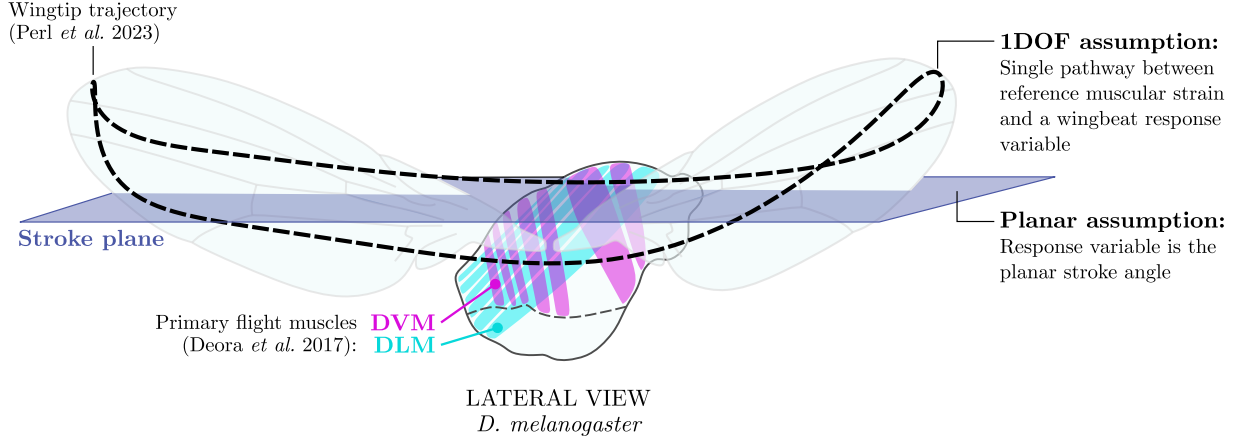

**Figure 1:** Schematic of the flight motor of *D. melanogaster*, based on [7,21], with 1DOF and planar oscillator modelling assumptions illustrated.

Finally, there is the broad question of model fidelity. The simplest conceivable oscillator model is the parallel-elastic linear oscillator [6,11]:

$$
\begin{aligned}
&M\ddot{x} + D\dot{x} + Kx = F, \\
&\ddot{x} + d\dot{x} + \omega_0^2 x = f, &&d = D/M, \ \ \omega_0^2 = K/M, \ \ f = F/M, \\
&\ddot{x} + 2\zeta\omega_0\dot{x} + \omega_0^2 x = f, &&\zeta = d/\omega_0,
\end{aligned}
\tag{1}
$$

with canonical parameters $\omega_0$, the structural natural frequency; and $\zeta$, the damping ratio. With appropriate definition of force ($F$, $f$), the kinematic variable $x$ can be taken either as the wingbeat stroke angle ($\phi$), muscular strain ($\varepsilon$), or, a normalised reference (*e.g.*, $-1 \leq x \leq 1$, for normative hovering flight). We note in passing that the formulation in damping ratio, $\zeta$, implies that dissipation scales with structural stiffness ($\omega_0$) which is unrealistic for aerodynamic dissipation. The formulation in damping time constant $d$ is likely more suitable for insect flight motors.

A key deficiency in the linear model (Eq. 1) is that wingbeat aerodynamic dissipation is certainly nonlinear [26]. A more realistic model is the nonlinear formulation of quadratic damping, *i.e.*, aerodynamic drag [5,26]:

$$
\begin{aligned}
&M\ddot{x} + C\dot{x}|\dot{x}| + Kx = F, \\
&\ddot{x} + \beta\dot{x}|\dot{x}| + \omega_0^2 x = f, &&\beta = C/M, \ \ \omega_0^2 = K/M, \ \ f = F/M,
\end{aligned}
\tag{2}
$$

with quadratic damping parameter $\beta$. To improve model fidelity further, exoskeletal structural damping [27,28], can be added. This requires a mixed time-frequency domain model:

$$
\begin{aligned}
&M\ddot{x} + C\dot{x}|\dot{x}| + F_{K,\gamma} = F, &&\hat{F}_{k,\gamma} = K(1 + i\gamma)\hat{x}, \\
& &&\mathcal{F}\{x\} = \hat{x}e^{i\Omega t}, \\
& &&\mathcal{F}\{F_{k,\gamma}\} = \hat{F}_{k,\gamma}e^{i\Omega t},
\end{aligned}
\tag{3}
$$

where $\gamma$ is the structural damping ratio, or loss tangent [29], and $\mathcal{F}\{\cdot\}$ denotes the Fourier transform. A mixed time-frequency formulation is required because $F_{K,\gamma}$ is *non-causal* [30,31]: in the time domain, $F_{K,\gamma}(t)$ depends on both past and future – and, as such, only approximates a real, causal, structure. More realistic models typically involve fractional-order terms of the form $K(i\Omega)^\gamma$, preserving causality and roughly constant damping with frequency [3,30].

More widely, frequency-domain analysis is a powerful tool for characterising motor self-oscillation. Eq. 1-3 permit frequency-domain representations in some form. Denoting $\mathcal{F}\{x\} = \hat{x}e^{i\Omega t}$ and $\mathcal{F}\{f\} = \hat{f}e^{i\Omega t}$, the linear model (Eq. 1) permits the exact representation:

$$(-\Omega^2 + 2i\zeta\Omega + \omega_0^2)\hat{x} = \hat{f}, \text{ or}$$
$$(-\Omega^2 + id\Omega + \omega_0^2)\hat{x} = \hat{f}. \tag{4}$$

The quadratic model (Eq. 2) requires approximation. If the response, $x(t)$, is simple-harmonic, $x = \hat{x}e^{i\Omega t}$, approximating observed *D. melanogaster* wingbeat kinematics [12,32], then an approximate representation is:

$$(-\Omega^2 + ic\Omega^2 + \omega_0^2)\hat{x} = \hat{f}, \tag{5}$$

where the quadratic damping ratio $c$ is dependent on amplitude $\hat{x}$. A derivation is given in the Supporting Information (§1), alongside the relations $c = \beta\hat{x}$ for a peak-damping approximation; $c = \pi\beta\hat{x}/4$ for an average-damping approximation. In general, the window $\pi/4 \leq c/\beta\hat{x} \leq 1$ is a suitable approximation of $\beta\hat{x}$ by $c$.

The quadratic damping ratio $c$ has several interesting properties. It is dimensionless: the ratio of peak forces arising from dissipative and inertial effects, both scaling with $\Omega^2$ [26]. As such, it is the inverse of the Weis-Fogh number ($N_{\mathrm{WF}}$) characterised by Lynch *et al.* [15]: the relation $c = 1/N_{\mathrm{WF}}$ allows a translation from reported Weis-Fogh numbers. Going further, the relationship between quadratic and linear damping ratios, $c$ and $\zeta$, is $2\zeta\omega_0 = c\Omega$. If $\omega_0 \approx \Omega$ – either, as a reference frequency or an assumed motor operating point [6] – then $c \approx 2\zeta$, allowing rough translation from reported motor linear damping ratios [6]. Finally, while Eq. 5 as presented looks like the Fourier transform of a linear time-domain model, we suspect it to be non-causal (*cf.* Eq. 3).

## 2.2. Modelling flight motor elasticity

In addition to differing models of dissipation, differing models of elasticity are also available. For one, both parallel and series elasticities might be present in the flight motor [15,33]. Introducing series elasticity to the quadratic model, Eq. 2, leads to the hybrid system:

$$M\ddot{x} + C\dot{x}|\dot{x}| + K_p x = F$$
$$M\ddot{x} + C\dot{x}|\dot{x}| + (K_p + K_s)x = K_s u \tag{6}$$
$$\text{with } F(t) = K_s(u - x),$$

where $K_p$ and $K_s$ are parallel and series stiffnesses, respectively, and $u$ is the muscular DOF, now distinct from $x$. Eq. 6 is expressible in the frequency domain. With canonical parameters $\omega_{0,p}^2 = K_p/M$, $\omega_{0,s}^2 = K_s/M$, $\beta = D/M$, and $f = F/M$ [6]; and the approximation window $\pi/4 \leq c/\beta\hat{x} \leq 1$:

$$(-\Omega^2 + ic\Omega^2 + \omega_{0,p}^2)\hat{x} = \hat{f},$$
$$(-\Omega^2 + ic\Omega^2 + \omega_{0,p}^2 + \omega_{0,s}^2)\hat{x} = \omega_{0,s}^2\hat{u}, \tag{7}$$
$$\text{with } \hat{f} = \omega_{0,s}^2(\hat{u} - \hat{x}).$$

Eq. 7 reduces to pure series elasticity for $\omega_{0,p} = 0$, and pure parallel elasticity for $\omega_{0,s} \to \infty$. To avoid the limit $\to \infty$, the stiffness $\omega_{0,s}$ may be reparametrized by a compliance parameter $\alpha$, for instance, as $\omega_{0,s} = \omega_{0,p}/\alpha$, *cf.* [6].

There is also the general question of elastic nonlinearity: thoracic exoskeletal structures from several insect species have been observed to show strain-hardening nonlinearity [27,34]. An analysis in the Supporting Information (§2), illustrates how a cubic-type elasticity can be accounted for in linear model of the form of Eq. 5, within an appropriate approximation window. As such, Eq. 5 can account for both cubic stiffness and quadratic damping: it is a highly general model, and will be the normative (but not exclusive) model used in this study.

### 2.3. Identifying motor model properties

The models of §§2.1-2.2 require parameter values: in particular, dissipative parameters $c$, $\beta$, and/or $d$, which we identify from the *D. melanogaster* meta-dataset of Pons *et al.* [12]. This dataset contains wingbeat load requirement profiles, meta-analysed from a range of reported experimental and computational data. For parameter identification, the inelastic loads [35] of the linear and quadratic aerodynamic dissipation models, Eq. 1-2, can be expressed:

$$G_D = M\ddot{x} + D\dot{x} = S_d(\ddot{x} + d\dot{x}),$$
$$G_C = M\ddot{x} + C\dot{x}|\dot{x}| = S_\beta(\ddot{x} + \beta\dot{x}|\dot{x}|),$$

(8)

where $S_d$ and $S_\beta$ are scale factors, and $-1 \leq x \leq 1$ for normative hovering flight. We consider the linear model to enable comparisons to the classical results of Machin and Pringle [4]. Using least-squares fitting we identify $\{S_d, d\}$ and $\{S_\beta, \beta\}$ for all load requirement profiles associated with biological wingbeat kinematics. Figure 2 illustrates these fits for the most heavily and most weakly damped profiles in the meta-dataset; in the Supporting Information (§4), fit parameters are given for every profile. Across the meta-dataset, $d \in [0.99, 1.76] \cdot 10^3$ rad/s, mean $d = 1.38 \cdot 10^3$ rad/s; and $\beta \in [0.90, 1.65]$, mean $\beta = 1.25$. Due to differing conditions across the meta-dataset, including in wing geometry and kinematics, the variation in damping values is relatively wide, spanning $\pm 35\%$ of the mean. However, the fit accuracy is good: $R^2 > 0.9$ for all profiles except those utilising the aerodynamics of Muijres *et al.* [36], which show $R^2$ down to 0.83 (Supporting Information, §4). Using the approximation window $\pi/4 \leq c/\beta\hat{x} \leq 1$, with $\hat{x} = 1$, we may estimate the quadratic damping ratio as $c \in [0.70, 1.65]$, mean $c = 1.11$, for normative hovering flight. For comparison, this is a Weis-Fogh number ranging over $N_{WF} \in [0.60, 1.42]$, or, if $\omega_0 = \Omega$ for the linear model, a linear damping ratio ranging over $\zeta \in [0.36, 0.64]$, mean $\zeta = 0.50$.

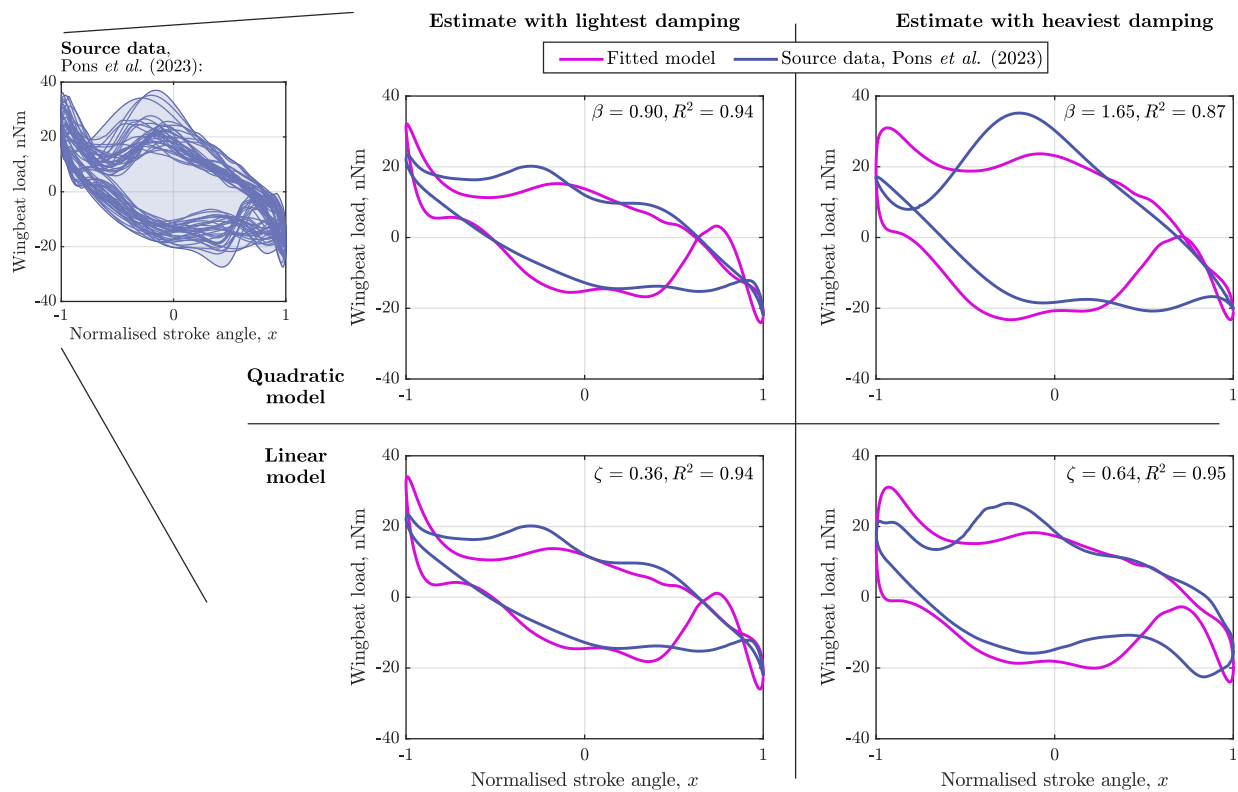

**Figure 2:** Motor models fitted to source data for *D. melanogaster* – quadratic ($\beta$) and linear ($d$, $\zeta$) models, illustrated for the most lightly damped ($\beta = 0.90$) and most heavily damped ($\beta = 1.65$) load requirement work loops in the source meta-dataset of Pons *et al.* [12]. The linear model is expressed in $\zeta$ rather than $d$ for ease of interpretation.

# 3. Modelling asynchronous muscle stretch-activation

## 3.1. Modelling asynchronous muscle dynamics

In the motor models of Eq. 1-7, the forcing, $F$ or $f$, is that of the flight muscles. In Dipterans, this includes both the primary flight muscles – the sets of asynchronous and antagonistic dorsoventral (DVM) and dorsolongitudinal (DLM) muscles (Fig. 1) – and an array of steering muscles. Neglecting steering muscles, which account for <3% of total muscle mass [37], we may view asynchronous muscle forcing through a viscoelastic lens. At low strain amplitudes, the muscle may be modelled as having a frequency-dependent complex modulus, $\mathbb{E}(\Omega)$: the ratio of muscular stress ($\sigma$) to strain ($\varepsilon$), with real-valued storage ($E$) and loss ($H$) moduli components [3,38]:

$$\mathbb{E}(\Omega) = \frac{\sigma(\Omega)}{\varepsilon(\Omega)} = E(\Omega) + iH(\Omega). \tag{9}$$

The storage modulus represents how elastic the muscle is (or, appears to be): the effective spring stiffness of the forces it generates. The loss modulus represents how much power it produces (or dissipates): its effective damping. When the loss modulus is negative, the muscle produces power, and does work on the motor to move the wings. With this in mind, we may express $\mathbb{E}(\Omega)$ in terms of the viscoelastic negative loss tangent ($r$) [29]:

$$\mathbb{E}(\Omega) = E(\Omega)\big(1 - ir(\Omega)\big), \qquad r(\Omega) = -\frac{H(\Omega)}{E(\Omega)}, \tag{10}$$

related also to the phase offset of muscular forcing, $\delta$, as $r = -\tan\delta$ [29].

Eq. 9-10 are the basis for a model of muscular forcing, $f(t)$. First, if the DVM and DLM operate as an exactly antagonistic pair (180° phase offset), then this pair resolves to a single equivalent $\mathbb{E}(\Omega)$. Then, in the frequency domain, $\hat{f}(\Omega) \propto \mathbb{E}(\Omega)\varepsilon(\Omega)$, where the constant of proportionality accounts for muscular cross-sectional area and an assumed linear kinematic relationship between muscular strain ($\varepsilon$) and the wingbeat kinematic variable ($x$ or $u$). As such:

$$\hat{f}(\Omega) = -N_E\big(E(\Omega) + iH(\Omega)\big)\hat{x} = -N_M\big(1 - ir(\Omega)\big)\hat{x}. \tag{11}$$

where $N_E$ and $N_M(\Omega)$ are undetermined constants of proportionality. These constants will be identified during the analysis; however, as detailed in the Supporting Information (§3), we may also estimate $N_E$ for *D. melanogaster* as:

$$N_E = \frac{A_{\text{muscle}} L_{\text{muscle}} \hat{\varepsilon}^2_{\text{muscle}}}{I_{\text{wing}} \hat{\phi}^2_{\text{wing}}} \approx 10.5 \text{ m kg}^{-1}, \tag{12}$$

given reported muscle length ($L_{\text{muscle}}$), total cross-sectional area ($A_{\text{muscle}}$), and strain amplitude ($\hat{\varepsilon}_{\text{muscle}}$); and wing inertia ($I_{\text{wing}}$) and stroke amplitude ($\hat{\phi}_{\text{wing}}$, $\pm$, in rad).

Specific forms of $\mathbb{E}(\Omega)$ for *D. melanogaster* are well-established. Two key forms are those of Kawai and Brandt [39], applied to insect flight muscle by Sicilia and Smith [2]; and the modified form, specific to insect flight muscle, used by Maughan *et al.* [3] and others [38,40]. Both these forms model three muscular processes, denoted A, B, and C, as:

Kawai-Brandt-type [2,39]:

$$\mathbb{E}(\Omega) = \underbrace{H + \frac{Ai\Omega}{a + i\Omega}}_{\text{Process A}} - \underbrace{\frac{Bi\Omega}{b + i\Omega}}_{\text{Process B}} + \underbrace{\frac{Ci\Omega}{c + i\Omega}}_{\text{Process C}} \, ,$$

Maughan-type [3,38,40]:

$$\mathbb{E}(\Omega) = \underbrace{A(i\Omega)^k}_{\text{Process A}} - \underbrace{\frac{Bi\Omega}{b + i\Omega}}_{\text{Process B}} + \underbrace{\frac{Ci\Omega}{c + i\Omega}}_{\text{Process C}} \, ,$$

(13)

where our $b$ and $c$ correspond to $2\pi b$ and $2\pi c$ in [38,40]. Process A represents the viscoelasticity of passive structures in the muscle [3,41]. The fractional-order term $A(i\Omega)^k$, $k \in [0, 1]$ preserves both causality and roughly constant damping with frequency, *cf.* [30,31]. Process B and C arise from muscular activation. The conversion of chemical energy (ATP) to mechanical work during activation occurs through the cross-bridge cycle [41] – the cyclic interaction between myosin- and actin-containing filaments – and this cycle is associated with changes in complex moduli. Process B is a work emitting process, and Process C a work absorbing process [3,38,39]. For both model types, process parameters can be identified in *ex vivo* empirical results for both steady and transient response testing. Indeed, muscle behaviour as per Eq. 13 can be seen as velocity feedback: a transfer function, $\mathbb{E}(\Omega)/i\Omega$, between stress and strain rate, identifiable from strain-rate impulse response data [42,43]. In addition, we note that, while nonlinearity in stretch-activated muscles is known to occur [44,45], connecting nonlinear model formulations [2,14] with observed behaviour is difficult.

Combining the muscle models of Eq. 9-13 with the motor models of §2.2 leads to frequency-domain characteristic equations for the motor – equations describing a free (unforced) system, in which self-oscillation may emerge. For instance, the quadratic model (Eq. 5) under general complex modulus excitation (Eq. 11), has the characteristic equation:

$$\begin{aligned} -\Omega^2 + ic\Omega^2 + \omega_0^2 &= -N_E\big(E(\Omega) + iH(\Omega)\big), \\ &= -N_M(\Omega)\big(1 - ir(\Omega)\big), \end{aligned}$$

(14)

a combination which we will use as a normative model. Analysing characteristic equations, such as Eq. 14, can proceed in a forward or an inverse direction: analysing the response ($\Omega$) for given structural properties ($c$, $\omega_0$, $N_E$, ...); or the structural properties required to effect a given response, respectively. We will pursue both analysis directions over §4.

### 3.2. Identifying muscle model properties

Parameters in Eq. 13 can be fitted to *ex vivo* muscular testing data [42–44]. In this study, we use data from two independent sources: firstly, and principally, the complete wild-type *D. melanogaster* sinusoidal testing dataset reported by Viswanathan *et al.* [46]. This dataset is composed of test data and fitted parameters for ten muscle fibre sinusoidal tests, carried out in solutions of suitably high-concentration ATP (the muscular energy source); and concentrations of calcium (the ion triggering activation) at which power output is maximised [38,46]. Fig. 3 illustrates this dataset, alongside the reported fitting to the Maughan-type model, in moduli and negative loss tangent. Reported fit parameter values range over $A \in [132, 373]$, $k \in [0.088, 0.112]$, $B \in [647, 4061]$, $b \in [1393, 2032]$, $C \in [651, 3921]$, and $c \in [2382, 3091]$. As can be seen in Fig. 3, moduli estimates are reasonably widely spread, varying by a factor of $2\times$. Estimates of the peak negative loss tangent range over 0.3 to 0.6, consistent with other studies of *D. melanogaster* indirect flight muscle (Table 1). The calculations behind the estimates of Table 1 are given in the Supporting Information (§5).

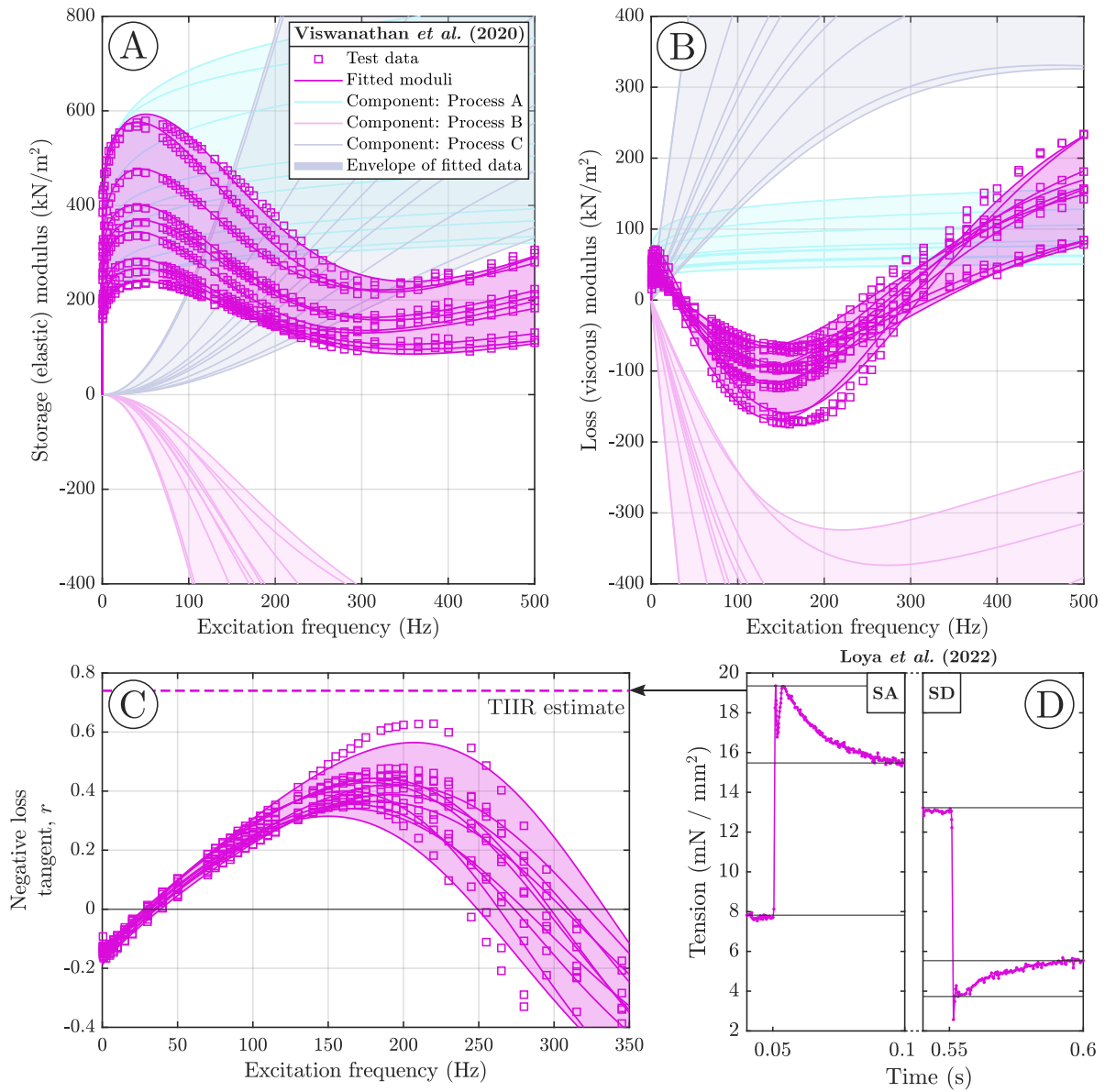

**Figure 3:** *Ex vivo* muscular data, leading to estimates of muscular negative loss tangent, $r$. Sinusoidal testing dataset from Viswanathan *et al.* [46], with associated fit, and process-wise composition (Processes A, B, C): (**A**) Storage modulus, (**B**) loss modulus, (**C**) negative loss tangent. (**D**) Strain rate impulse response data for stretch-activation (SA) and shortening-deactivation (SD) from Loya *et al.* [42], with associated TIIR estimate of $r$ overlaid on (**C**). This TIIR estimate is the largest among the impulse-response data for *D. melanogaster* surveyed in Table 1.

Secondly, we utilise the impulse-response characterisation of the dual phenomenon of stretch-activation (SA) and shortening-deactivation (SD, *i.e.*, SA upon contraction) in wild-type *D. melanogaster* flight muscle reported by Loya *et al.* [42]. It is possible to fit Eq. 13 directly to this impulse-response data, *cf.* [5], but raw data and the full set of fitted rate constants ($b, c, \ldots$) are rarely reported [42]; and amplitude constants ($B, C, \ldots$) are typically unreported. For this reason, we develop our own time-independent impulse-response (TIIR) estimates of the peak $r$, based on only on a few measured SA/SD measurements – accounting for the directionally-nonlinear distinction between SA and SD. Details of these estimates are given in the Supporting Information (§4). The TIIR estimate for the single pair of SA/SD profiles reported by Loya *et al.* [42], is peak $r \approx 0.74$ (Fig. 3C, Table 1). The TIIR process is also applicable to other impulse-response datasets in the literature: the Supporting Information (§6) presents TIIR calculations for the SA-only data of Glasheen *et al.* [47], under two different muscle bathing solutions, and of Wang *et al.* [48], under two different strain step changes. Estimated peak $r$ values for these datasets are shown in Table 1. There is broad consistency between estimates of $r$ across sinusoidal testing and impulse response datasets, with no estimate exceeding $r \approx 0.74$ [42]. We note, however, that these estimates are coarse: in particular, TIIR estimates are likely to represent overestimates of the peak $r$ for their respective datasets; but also, do not account for nonlinearities that may make muscular oscillation differ from step response behaviour [49]. The value of the TIIR process is that it allows conservative estimates of maximum $r$ – a crucial parameter for our analysis over §§4-5 – to be made using only low-precision graphs of strain rate impulse response, which are unsuitable for detailed fitting to viscoelastic models such as Eq. 13.

**Table 1:** Estimates of peak $r$ for *D. melanogaster* indirect flight muscle.

| Dataset | Estimated peak $r$ |
|---|---|
| **Sinusoidal testing:** | |
| Trujillo *et al.* (2021) [50] | 0.48 |
| Yang *et al.* (2010) [41] | 0.32 |
| Barton *et al.* (2005) [51] | 0.16 |
| Henkin *et al.* (2004) [52] | 0.31 |
| Dickinson *et al.* (1997) [40] | 0.38 |
| Warmke *et al.* (1992) [53] | 0.55 |
| **Strain rate impulse response (TIIR):** | |
| Loya *et al.* (2022) [42] | 0.74 |
| Glasheen *et al.* (2017) [47] | |
|     Swank solution | 0.46 |
|     Reedy solution | 0.26 |
| Wang *et al.* (2011) [48] | |
|     1% strain step change | 0.46 |
|     2.5% strain step change | 0.39 |

In general, these peak $r$ estimates are for muscular strain amplitudes ($\hat{\varepsilon}$) lower than those anticipated *in vivo*. In *D. virilis* tethered flight, these amplitudes average ±1.8% [54], though these estimates are coarse, and apparently inconsistent with the observation that *D. melanogaster* flight muscle generates maximum power at amplitudes as low as ±0.4% [48]. Nevertheless, Viswanathan *et al.* [46] prescribe an amplitude of only ±0.0625%; Loya *et al.* [42], an effective ±0.5%. These low strain amplitudes are a key limitation of this study: both because a complex-modulus model may break down at high amplitudes [44,45], and because the parameters of any model may change between low and high amplitudes. Model breakdown can be overcome, to a degree, via direct analysis of work loop data (§4.2); but parameter variation remains unclear. Only a TIIR estimate derived from Wang *et al.* [48] reaches clearly biological levels of ±1.3%, and while this estimate is consistent (peak $r \approx 0.39$), nonlinearities with respect to strain amplitude could still lead to qualitatively different behaviour at high strain amplitudes.

# 4. Conditions for motor self-oscillation

## 4.1. Self-oscillation conditions for PEA motor systems

We begin by considering a forward-problem approach to the normative *D. melanogaster* motor model, Eq. 14. In this initial analysis, we take the stiffness parameter ($N_M$) and negative loss tangent ($r$) to be given constants – in theory, their values at the motor self-oscillation frequency, which is as yet undetermined. In practice, $r$ is confined to a narrow window: between 0 and 0.6 for *D. melanogaster* motor operation at any frequency (Fig. 3). $N_M$, we take as unknown. As such, the motor characteristic equation can be expressed:

$$-\Omega^2 + ic\Omega^2 + \omega_0^2 = -N_M + iN_Mr, \tag{15}$$

and solved for the complex-valued operating characteristic, $\Omega$. In general, $\Omega$ has two solutions: $\Omega_\pm = \mathfrak{Re}(\Omega_\pm) + i\mathfrak{Im}(\Omega_\pm)$, only one of which is physical. The real part, $\mathfrak{Re}(\Omega_\pm)$, is the response frequency – only the solution branch with positive real part, denoted $\Omega_+$, is physical. The imaginary part, $\mathfrak{Im}(\Omega_+)$, is the response dissipation factor. $\mathfrak{Im}(\Omega_+) < 0$ represents divergent oscillation; $\mathfrak{Im}(\Omega_+) > 0$ convergent oscillation; and $\mathfrak{Im}(\Omega_+) = 0$ critical self-oscillation: the steady operation of an insect flight motor.

Eq. 15's solution in $\Omega$ is:

$$\Omega_\pm = \pm\sqrt{\frac{iN_Mr - N_M - \omega_0^2}{ic - 1}}, \tag{16}$$

which can be reformulated as:

$$\Omega_\pm = \mathfrak{Re}(\Omega_\pm) + i\mathfrak{Im}(\Omega_\pm),$$

with:

$$\mathfrak{Re}(\Omega_\pm) = \pm\sqrt{\frac{\Delta + N_M(1 + cr) + \omega_0^2}{2(c^2 + 1)}},$$

$$\mathfrak{Im}(\Omega_\pm) = \pm\,\text{sgn}(N(c - r) + c\omega_0^2)\sqrt{\frac{\Delta - N_M(1 + cr) - \omega_0^2}{2(c^2 + 1)}}, \tag{17}$$

$$\text{where } \Delta = \sqrt{(c^2 + 1)(N_M^2(1 + r^2) + 2N_M\omega_0^2 + \omega_0^4)}.$$

These expressions are derived with symbolic computation in MATLAB-MuPAD [55] – see the Data Availability Statement.

Eq. 17 gives the conditions for self-oscillation to manifest. Locating $\mathfrak{Im}(\Omega_+) = 0$, we compute a condition for critical self-oscillation in terms of a critical quadratic damping factor, $c = c_{\text{crit}}$:

$$c_{\text{crit}} = \frac{N_Mr}{N_M + \omega_0^2} \quad \text{for critical self-oscillation.} \tag{18}$$

We may confirm with symbolic computation that, under any $N_M$, $r$ and $\omega_0 > 0$, if $c < c_{\text{crit}}$ the system will diverge, and if $c > c_{\text{crit}}$, it will converge. If then, we have a motor with given peak $r$ (a muscle of fixed properties) but with $N_M$ and $\omega_0$ that we could hypothetically control, then the maximum possible $c$ that we could self-excite would be:

$$c_{\text{max}} = r. \tag{19}$$

That is to say, it is not possible to self-excite a system with $c > r$ under *any* conditions – no matter the exoskeletal stiffness/elasticity ($\omega_0$) *or the quantity of muscle* ($N_M$). This is a remarkable limit on the kinds of flight motor for which self-oscillation is possible. And it is a limit that forms the beginnings of a paradox – given that reported $r$ values (0-0.6, §3.2) and $c$-values (0.9-1.65, §2.3) apparently do not obey this limit.

### 4.2. The self-oscillation paradox in *D. melanogaster*

To generalise the analysis of §4.1 to frequency-dependent moduli, we transition to an inverse-problem analysis. If the motor is specified to self-oscillate steadily at a given frequency $\Omega = \Omega_*$, then the characteristic equation for operation of our normative motor model at this state is:

$$-\Omega_*^2 + ic\Omega_*^2 + \omega_0^2 = -N_E\big(E(\Omega_*) + iH(\Omega_*)\big), \tag{20}$$

where $E(\Omega_*)$ and $H(\Omega_*)$ are given by the moduli models in §3.2. The unknown variables are the level of dissipation ($c$) and exoskeletal elasticity ($\omega_0$) in the motor, and the muscle factor ($N_E$) required for steady self-oscillation. If we are interested in the maximum dissipation ($c_{\max}$) at which the motor is capable of self-oscillation, then, we may solve this inverse problem for the critical muscular factor ($N_{E,\text{crit}}$) and dissipation ($c_{\text{crit}}$) consistent with self-oscillation:

$$-\Omega_*^2 + ic_{\text{crit}}\Omega_*^2 + \omega_0^2 = -N_{E,\text{crit}}\big(E(\Omega_*) + iH(\Omega_*)\big),$$
or:
$$\Omega_*^2 - \omega_0^2 = N_{E,\text{crit}}E(\Omega_*),$$
$$c_{\text{crit}}\Omega_*^2 = -N_{E,\text{crit}}H(\Omega_*), \tag{21}$$

and thus:

$$N_{E,\text{crit}} = -\frac{c_{\text{crit}}\Omega_*^2}{H(\Omega_*)} = \frac{\Omega_*^2}{E(\Omega_*)}\left(1 - \frac{\omega_0^2}{\Omega_*^2}\right), \quad c_{\text{crit}} = r(\Omega_*)\left(1 - \frac{\omega_0^2}{\Omega_*^2}\right). \tag{22}$$

Eq. 22 has three implications. *First*, the motor never oscillates at below the exoskeletal natural frequency, $\omega_0$: the presence of muscular elasticity, $N_{E,\text{crit}}E(\Omega_*)$, necessitates that $\Omega_* > \omega_0$. *Second*, exoskeletal elasticity does not reduce the burden on the muscles: the quantity of muscle required to self-oscillate a given level of damping, measured as $N_{E,\text{crit}}/c_{\text{crit}}$, is constant w.r.t. $\omega_0$. *Third*, consistent with §4.1, $c_{\text{crit}}$ decreases with exoskeletal elasticity $\omega_0$, and reaches maximum $c_{\max} = r(\Omega_*)$ at $\omega_0 = 0$. As such, $c_{\max} = r(\Omega_*)$ is a hard limit on the motor's capability for self-oscillation. If we track $c_{\text{crit}}$ over $\Omega_*$ for the muscular data of Viswanathan *et al.* [46], we observe that it is impossible for *D. melanogaster* flight muscle to excite a motor with dissipation greater than $c \approx 0.6$ (Fig. 4A); or, in the best-case TIIR estimate, $c \approx 0.74$.

This limit presents a paradox: by several estimates, the level of dissipation in *D. melanogaster* during normative hovering flight exceeds this limit significantly. Fig. 4 illustrates this paradox in three different ways: in the inconsistency between the maximum quadratic damping that the muscle and exoskeletal transmission can cause to self-oscillate, and the actual quadratic damping model for the wingbeat (Fig. 4A); in the analogous inconsistency for linear damping (Fig. 4B); and in a direct inconsistency of work loops for muscular forcing and wingbeat load requirements (Fig. 4D). The level of elasticity, *i.e.*, level of tilt, in these loops is inconsistent, and cannot be made consistent by any scale factor ($N_E$, $N_H$ – these do not alter the ellipse tilt); nor by any addition of exoskeletal elasticity ($\omega_0$ – this exacerbates the inconsistency). Forcing and loading are fundamentally inconsistent.

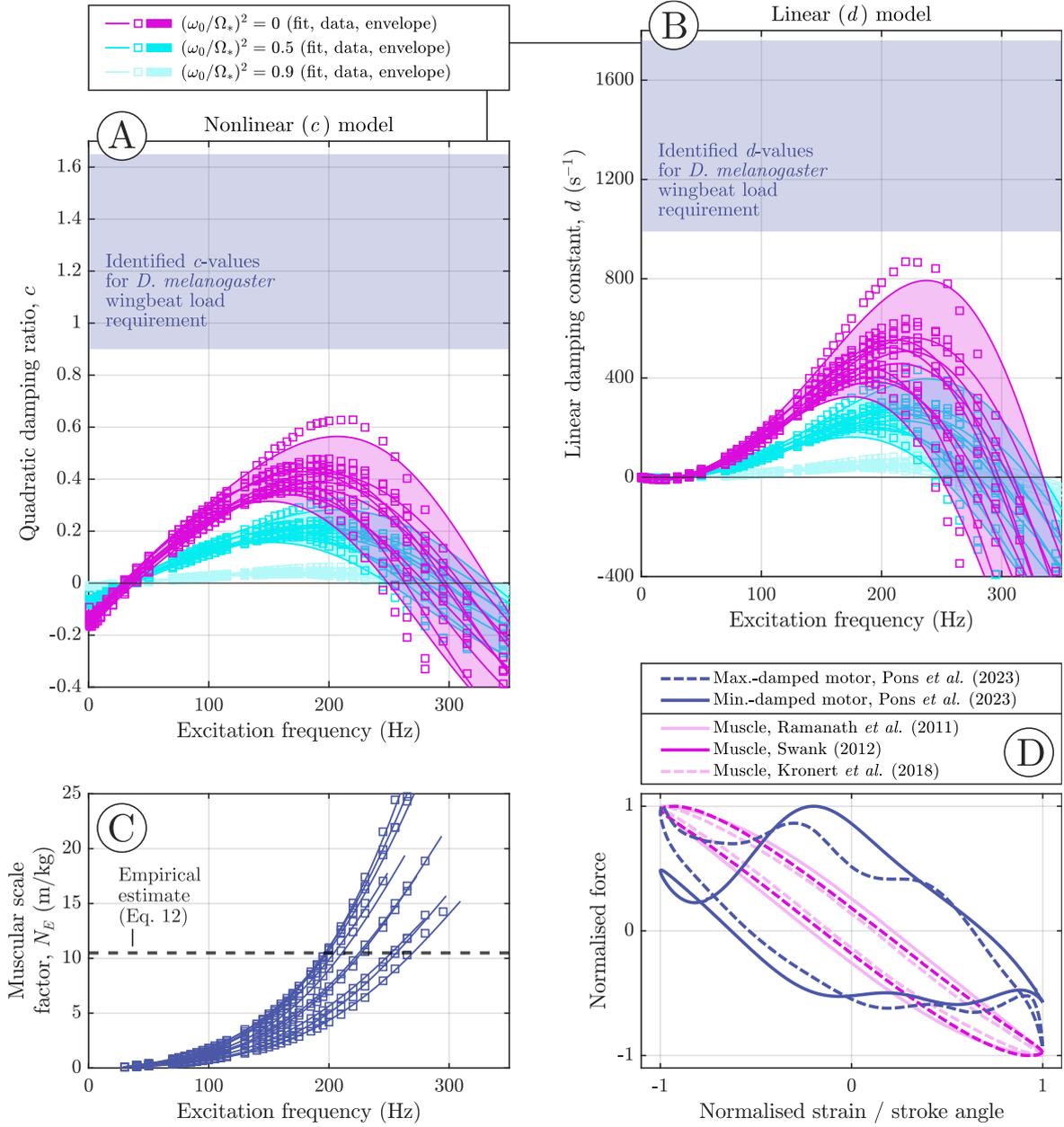

**Figure 4:** The self-oscillation paradox in the flight motor of *D. melanogaster*. (**A**) Critical levels of damping for the nonlinear model, $c_{\text{crit}}$, as a function of the exoskeletal stiffness, $(\omega_0/\Omega_*)^2$, and compared to estimates of $c$ for the flight motor (Fig. 2). (**B**) Critical levels of damping for the linear model, $d_{\text{crit}}$, as a function of the exoskeletal stiffness, compared to estimates of motor $d$ (Fig. 2). (**C**) The critical muscular scale factor, $N_{E,\text{crit}}$, required for self-oscillation under $\omega_0/\Omega_* = 0$, compared with the estimate of $N_E$ from motor properties (Eq. 12). Note that the ratios $N_{E,\text{crit}}/c_{\text{crit}}$ and $N_{E,\text{crit}}/d_{\text{crit}}$ are independent of $\omega_0/\Omega_*$. (**D**) The self-oscillation paradox illustrated in empirical nonlinear data: work loops for the motor [12] and muscle [38,44,56], illustrating the difference in relative damping.

### 4.3. The self-oscillation paradox across motor models

One question of immediate relevance is whether this paradox holds across different motor models – representing, different physical configurations and effects that may be present within the motor. Is it simply a feature of our selected model? We show that it is not, by applying the same symbolic computation-based forward and/or inverse analyses as in §§4.1-4.2 to the differing models of motor dissipation and elasticity described in §§2.1-2.2. The details of this process are given in the Supporting Information (§7) – to summarise:

*First*, the addition of exoskeletal structural damping (Eq. 3) reduces the reduce the level of aerodynamic damping that is permissible while maintaining self-oscillation. Exoskeletal damping, certainly present at some level within the *D. melanogaster* flight motor, makes self-oscillation even more difficult.

*Second*, the scaling of aerodynamic damping in the motor may not be exactly quadratic, given the low Reynolds number of operation (~100) [12]. As an example of lower-degree scaling, if damping is taken to be linear ($d$, Eq. 1), then, under equivalent forward and inverse formulations:

$$\text{Forward:} \quad d_{\text{crit}} = \frac{N_M r}{\sqrt{N_M + \omega_0^2}}, \qquad d_{\text{max}} = r\sqrt{N_M},$$

$$\text{Inverse:} \quad d_{\text{crit}} = r(\Omega_*)\Omega_* \left(1 - \frac{\omega_0^2}{\Omega_*^2}\right), \quad d_{\text{max}} = r(\Omega_*)\Omega_*. \tag{23}$$

In contrast with $c_{\text{max}}$, $d_{\text{max}}$ *does* scale with muscle mass $N_M$ – though this entails an increase in self-oscillation frequency, $\Re(\Omega)$ or $\Omega_*$. In theory, if $r > 0$ can be maintained, then it is always possible to out-scale linear damping with muscular forcing. However, apart from the relatively unrealistic scaling of the linear model, *D. melanogaster* flight muscle shows $r > 0$ only up to roughly 300 Hz (Fig. 3). Over this window, estimates of $d$ for *D. melanogaster* again exceed $d_{\text{max}}$ significantly (Fig. 4B).

*Third*, the distribution of elasticity across the flight motor may not be exactly parallel. Series elasticity (Eq. 6-7) can arise from the flexibility of the wing hinge [15,33] and local modes of deformation within the exoskeleton [6]; and could act as a first-order approximation of the elastic effects of distributed wing flexibility – though the deformation observed in *D. melanogaster* wings during flight is relatively minimal [32,57]. We can account for these effects via a hybrid parallel-series model of the flight motor (Eq. 7). Such a model has critical quadratic damping:

$$c_{\text{crit}} = \frac{r}{(1 + \alpha\nu)(1 + \alpha + \alpha r^2) + \nu(1 + \alpha)}, \qquad \nu = \frac{\omega_{0,p}^2}{N_M}, \tag{24}$$

under a forward analysis. For reference, a plot of this function is given in the Supporting Information (§7). From Eq. 24, positive $\alpha$ and $\nu$ can only reduce $c_{\text{crit}}$, and $c_{\text{max}} = r$. By the same token, linearised strain-hardening effects in any elastic term do not increase $c_{\text{max}}$, as $c_{\text{max}}$ occurs when exoskeletal elasticity is absent ($\nu = \alpha = 0$). Wing hinge flexibility and other series-elastic effects cannot alter the $c_{\text{max}} = r$ self-oscillation limit – a limit which now appears fundamental to a range of linearized oscillators.

*Fourth* – there is only one form of linearised elasticity that could resolve the self-oscillation paradox. Unstable elasticity, $\omega_0^2 < 0$, as could arise within the well of a bistable mechanism within the thoracic transmission, could enable $c_{\text{crit}} > r$. Estimates of the factor by which observed *D. melanogaster* damping ($c_{\text{obs}}$) exceeds peak muscular capability ($r_{\text{obs}}$, from Viswanathan *et al.* [46]) range from $c_{\text{obs}}/r_{\text{obs}} = 1.4$ in the best case, to 4.7 in the worst case (Fig. 4A). The levels of parallel instability (in Eq. 18, 22) generating these factors are:

$$\text{Relative instability} = -\frac{\omega_0^2}{N_M} = 1 - \frac{r_{\text{peak}}}{c_{\text{ID}}} \,, \tag{25}$$

a measure which varies from 0.30-0.79 across best-worst cases. Exoskeletal instability would have to be at least 30% as strong as effective elasticity of the muscles themselves to resolve the paradox. The only way to resolve the paradox simply by choice of model is to invoke instability, which is problematic in morphological terms, as we discuss further in §6.2.

## 5. Is motor self-oscillation resonant?

### 5.1. Excitation to effective energy resonance

In §4 we focused on the conditions for motor self-oscillation to emerge. A consequent topic concerns the frequency to which this self-oscillation converges, and how this frequency maps on to the resonant landscape of the motor [6]: its classical and/or energy resonant frequencies. Within the response frequencies, $\mathfrak{Re}(\Omega_+)$, of the motor models in §4, one central principle can be identified. The storage modulus, $E(\Omega)$, functions in a dynamical sense like elasticity – storing and releasing energy, or appearing to do so. If we treat $E(\Omega)$ as elasticity, we can divide any motor-muscle system into an *effective* system composed of the motor dynamics plus $E(\Omega)$; and an effective forcing, $H(\Omega)$. For any linearised motor dynamics $\mathbb{D}(\Omega)$:

$$\mathbb{D}(\Omega)\hat{x} = -N_E\big(E(\Omega) + iH(\Omega)\big)\hat{x},$$
$$\Rightarrow \underbrace{\big(\mathbb{D}(\Omega) + N_E E(\Omega)\big)\hat{x}}_{\substack{\text{effective} \\ \text{dynamics}}} = \underbrace{iN_E H(\Omega)\hat{x}}_{\substack{\text{effective} \\ \text{forcing}}} \,. \tag{26}$$

Then, if self-oscillation in the motor is simple-harmonic (§2), then it *necessarily* converges to the energy resonant frequency [6] of the effective dynamics, denoted $\omega_{\text{e,eff}}$. This is because the effective forcing $\hat{f}_{\text{eff}} = iN_E H(\Omega)\hat{x}$ is incapable of generating negative work and therefore can only generate effective energy resonant states:

$$\hat{f}_{\text{eff}} = iN_E H(\Omega)\hat{x},$$
$$f_{\text{eff}}(t) = \mathfrak{Re}\big\{\hat{f}_{\text{eff}}e^{i\Omega t}\big\} = -N_E H(\Omega)\hat{x}\sin\Omega t\,,$$
$$p_{\text{eff}}(t) = \dot{x}(t)f_{\text{eff}}(t) = N_E H(\Omega)\Omega\hat{x}^2\sin^2\Omega t \Rightarrow p_{\text{eff}}(t) \geq 0 \;\forall t. \tag{27}$$

If the effective dynamics are parallel-elastic, with linear, quadratic or viscoelastic damping, the effective energy resonant frequency is the effective natural frequency – the natural frequency associated with the elasticity of the exoskeleton plus activated muscle, *cf.* [6]. In the hybrid systems, provided self-oscillation is simple-harmonic, this frequency will be the effective energy resonant frequency, computable from [6][†].

### 5.2. Recovering and extending the results of Machin and Pringle [4]

Excitation to effective energy resonance provides theoretical context to the classical *ex vivo* experimental results of Machin and Pringle [4]: observations of self-oscillation in an asynchronous basalar muscle from the Indian rhinoceros beetle *Oryctes rhinoceros* connected to an artificial parallel-elastic linear oscillator. We compare the observations in Cases A-D of Fig. 8 in Machin and Pringle [4], to predictions in our modelling context – both as model validation, and to investigate the ways in which these *ex vivo* observations may break down *in vivo*. For each case, we extract reported muscular work loop data and estimate $r$, as detailed in the Supporting Information (§8). Computing also the artificial oscillator damping ratio, $\zeta =$ $D/2\sqrt{MK}$, from reported mass ($M$), stiffness ($K$), and damping coefficients ($D$), we predict the self-oscillation frequency $\Omega$ from the principle of excitation to effective energy resonance. By substituting $\mathbb{D}(\Omega) = -\Omega^2 + 2i\zeta\Omega + \omega_0^2$ (Eq. 4) and the $N_M(\Omega)$ forcing formulation (Eq. 11) into Eq. 26, we may solve for $\Omega = \omega_{\text{e,eff}}$, with $\omega_{\text{e,eff}}$ being the natural frequency of combined exoskeletal ($\omega_0^2$) and muscular ($N_M$) effects. That is:

$$\Omega^2 = \omega_{\text{e,eff}}^2 = \omega_0^2 + N_M^2(\Omega),$$

$$\frac{\Omega}{\omega_0} = \sqrt{1 + \frac{N_M^2(\Omega)}{\omega_0^2}} = \sqrt{1 + \frac{4\zeta^2\Omega^2}{r^2}} \quad \therefore \quad \frac{\Omega}{\omega_0} = \frac{\zeta}{r} + \sqrt{\left(\frac{\zeta}{r}\right)^2 + 1}, \tag{28}$$

using the estimates of $r$ from the recorded muscular work loops. The predictions of Eq. 28 match well the self-oscillation frequency actually observed by Machin and Pringle [4] (Fig. 5A-B). The errors observed (>15%) are explicable by factors such as uncertainty and non-ellipticity in the muscular work loops (Fig. 5C); and unmeasured damping phenomena (*e.g.*, aerodynamic damping). The comparison in Fig. 5B validates our self-oscillation analysis against experimental data for linear oscillators excited by insect flight muscle.

---

[†] There remain open questions in such systems – *e.g.*, does the non-existence of an energy-resonant frequency in certain hybrid systems [6] imply the impossibility of self-oscillation?

However, Machin and Pringle's [4] *ex vivo* data do not extend to the levels of damping associated with *D. melanogaster* hovering flight (Fig. 5A). We can perform this extension with the model: to $\zeta \approx 0.6$; a hypothetical $r = 1.6$ resolving the self-oscillation paradox; and separately, a quadratic damping model. The extension reveals ways in which these *ex vivo* results might not generalise to *D. melanogaster in vivo*. Two factors cause a breakdown: the level of damping, and its scaling with frequency. As damping increases, self-oscillation occurs at ascending frequencies – at hovering flight conditions, substantially higher than the exoskeletal natural frequency, $\omega_0$ (Fig. 5A). The nature of this frequency ascent is, however, dependent on how damping scales with frequency (Fig. 5D). This illustrates, following Ellington [26], the importance of precise models of frequency-based scaling of flight motor aerodynamic power consumption. The first key implication of Fig. 5A,D is that muscular self-oscillation may drive the system far from the exoskeletal natural frequency, $\omega_0$. In conditions representative of *D. melanogaster*, the frequency of self-oscillation is determined largely by muscular, not exoskeletal, effects.

Alongside this extension to *in vivo* conditions, we may also map the behaviour of other resonant frequencies within the motor effective dynamics. Following [6], we focus on displacement ($\omega_{\text{disp,eff}}$) and velocity ($\omega_{\text{vel,eff}}$) resonant frequencies:

$$\omega_{\text{disp,eff}} = \arg\max_{\Omega}\left(\frac{|\hat{x}|}{|\hat{f}_{\text{eff}}|}\right), \qquad \omega_{\text{vel,eff}} = \arg\max_{\Omega}\left(\frac{|i\hat{x}\Omega|}{|\hat{f}_{\text{eff}}|}\right). \tag{29}$$

For the linear and quadratic model, these frequencies are:

Linear:

$$\frac{\omega_{\text{disp,eff}}}{\omega_0} = \sqrt{1 - 2\zeta^2 + \frac{N_M}{\omega_0^2}}, \qquad \frac{\omega_{\text{vel,eff}}}{\omega_0} = \sqrt{1 + \frac{N_M}{\omega_0^2}},$$

Quadratic:

$$\frac{\omega_{\text{disp,eff}}}{\omega_0} = \sqrt{\frac{1 + \frac{N_M}{\omega_0^2}}{1 + c^2}}, \qquad \frac{\omega_{\text{vel,eff}}}{\omega_0} = \sqrt{\frac{1 + \frac{N_M}{\omega_0^2}}{\sqrt{1 + c^2}}},$$

$$\tag{30}$$

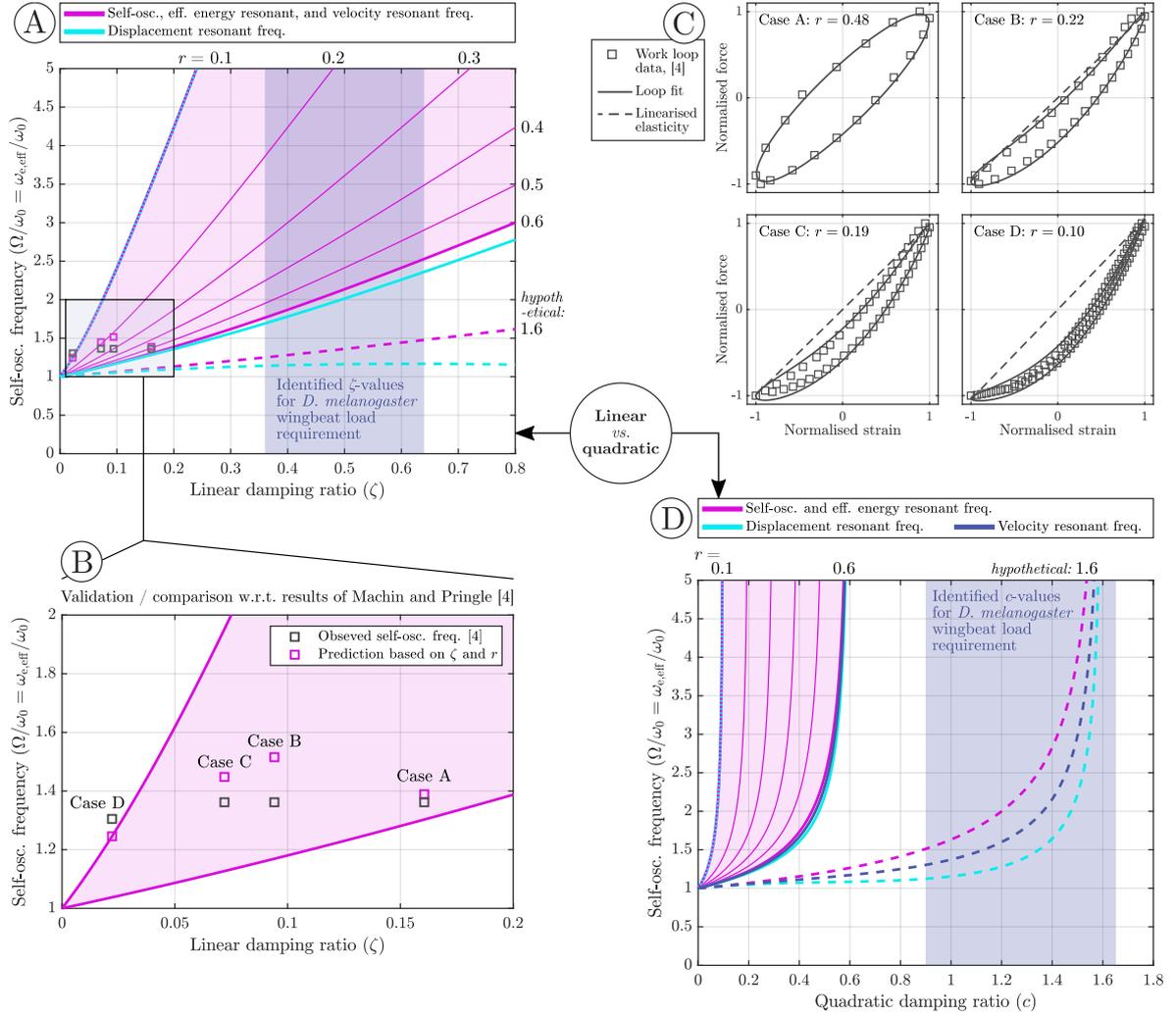

**Figure 5:** Relationship between self-oscillation frequency and effective resonant frequencies in linear and quadratic motor models, alongside the *ex vivo* experimental results of Machin and Pringle [4]. (**A**) Predictions of self-oscillation frequency for the linear model, alongside motor effective resonant frequencies; identified $\zeta$ for *D. melanogaster*; and Machin and Pringle's results. (**B**) Close-up of the comparison with Machin and Pringle's results, with Cases A-D indicated. (**C**) Fit results for the work loops recorded by Machin and Pringle, used to identify $r$ for model prediction. (**D**) Predictions of self-oscillation frequency for the quadratic model, alongside motor effective resonant frequencies.

with:

Linear: Quadratic:

$$\frac{N_M}{\omega_0^2} = \frac{2\zeta^2 + 2\zeta\sqrt{r^2 + \zeta^2}}{r^2}, \qquad \frac{N_M}{\omega_0^2} = \frac{c}{r - c}. \qquad (31)$$

As illustrated in Fig. 5, under Machin and Pringle's [4] conditions, all these frequencies are closely aligned, and motor self-oscillation can be considered resonant in multiple senses. However, under the heavy quadratic damping more representative of *D. melanogaster in vivo*, resonant frequencies can be widely separated, and motor self-oscillation can show substantial ($> 50\%$) deviation from classical resonance. The second key implication of Fig. 5A,D is that the conclusions of Machin and Pringle [4] break down *in vivo*: the *D. melanogaster* is not necessarily resonant in all senses of the term. It is likely energy resonant in the sense of effective energy resonance (§5.1) but is likely not resonant in the classical sense of peak transfer function response. We should note, however, that this result has two key limitations, discussed further in §6. *First*, the effective dynamics framework, Eq. 26, is an optimistic characterisation of the muscle's elastic behaviour: its real elastic energy storage may differ [12], and effective energy resonance may not translate to real energy savings. *Second*, following Machin and Pringle [4], we assume that stable exoskeletal elasticity ($\omega_0 > 0$) is present, and that the self-oscillation paradox is resolved via a muscle with high negative loss tangent ($r \approx 1.6$) – that is, a muscle with high power output and low stiffness. The alternative case of near-zero or unstable exoskeletal elasticity ($\omega_0 \leq 0$) is not covered by this treatment, and may lead to different conclusions regarding the nature of resonance in the motor. As per §6, further data-driven analysis is needed to resolve the self-oscillation paradox in the Dipteran flight motor before the nature of resonance in this motor can be precisely characterised.

# 6. Discussion and conclusion

## 6.1. Resolution of the self-oscillation paradox via muscular nonlinearity

The self-oscillation paradox is based on the inconsistency between observed muscular negative loss tangents, $r$, and reported wingbeat damping – quadratic damping ratio $c$, or other damping metrics. One resolution to the paradox is that either of these parameters are not representative of *D. melanogaster in vivo*. It is difficult to make a case that wing inertial forces are substantially greater, or wing drag forces smaller, than existing estimates [6,12,15] – the only realistic contributing factor in this direction is the possibility that distributed wing flexibility, which is not accounted for in the source aerodynamic models, reduces the drag load on the wing, and thus decreases $c$. To resolve the self-oscillation paradox *D. melanogaster*, wing flexibility would have to reduce drag forces ($c\Omega^2$) to below half of current estimates, maintaining roughly equivalent lift. The flexibility observed in *D. melanogaster* wings during flight is not large [32,57]; and existing studies of insectoid flexible flapping wings indicate reductions in drag-per-lift of up to ~20% in the very best cases [58–60]. As such, wing flexibility could contribute slightly to resolving the self-oscillation paradox in *D. melanogaster*, but does not offer a complete resolution.

However, there is a stronger case that estimates of $r$ may be unrepresentative. It is possible that there are biochemical factors that cause the muscle to generate substantially more power *in vivo*, at the same stiffness; or the same power at lower stiffness due to differences in activation conditions – *e.g.*, changes in role of elastic proteins [16] and/or the behaviour of the cross-bridge cycle [41]. It is also possible that active muscle is subject to strain-softening nonlinearity – decreasing stiffness with increasing strain – as observed in relaxed flight muscle [17]. Further data is needed on the nonlinear behaviour of activated muscle with respect to strain, as well as the actual *in vivo* strain amplitude of *D. melanogaster in vivo*. However, it should be noted that, while strain-based nonlinearities in general are known to be present [2,14,44,45], the very limited data available at high strain amplitude – a single TIIR estimate (Table 1) – would suggest that significant increases in $r$ are not occurring.

## 6.2. Resolution of the self-oscillation paradox via motor static instability

A second potential resolution concerns the fact that the condition $c \leq r$ for self-oscillation does not hold for exoskeletal unstable elasticity, or static instability. Contemplating exoskeletal instability recalls the outdated hypothesis of the Dipteran click mechanism [61,62]; but an instability to resolve the self-oscillation paradox could be qualitatively different. It could be relatively weak: at minimum only 30% as strong as muscular effective elasticity (§5.2). As such, it might not generate observable static instability in the motor even when the muscles are inactive. It would also be required only at small scales: not necessarily in larger Dipterans, and perhaps in small insects from other orders, such as Ptilids [63]. Indeed, there is a general question of the scaling and generality of the self-oscillation paradox. If, as expected, the paradox resolves naturally ($c \leq r$) at larger scales of insect, what is the threshold? And is the same paradox present in other insects at similar or smaller scales than *D. melanogaster*? Ptilids are a particularly relevant test case, as the bristled wings of these tiny beetles generate very low inertial forces – as low as 1% of aerodynamic forces [64]. This leads to an extremely dissipation-dominated wingbeat [63], *i.e.*, $c \to \infty$, which could only emerge under a flight muscle with *no* effective stiffness ($r \to \infty$), or under exoskeletal static instability. Differences between Ptilid and Dipteran muscle architecture could be a confounding factor [65], but it appears insufficient to postulate a mechanism that only reduces the elasticity/stiffness of the motor system: this elasticity must be brought to zero.

There is also a further point here, alluding to §7.1. A different static instability is already present in the motor: the muscle cross-bridge dynamics. A notable feature of the A-B-C decomposition of muscular processes (§3.1) is that the cross-bridge cycle, processes B and C, acts as static instability: reducing the effective elasticity of the muscle (Fig. 3). In reported data, the cross-bridge dynamics are not strong enough to drive the muscle into static neutral stability or instability, but if this effect was more powerful *in vivo*, or muscular structural elasticity (process A) was weaker, this might occur. Cross-bridge dynamics are a further potentially-nonlinear phenomenon to track with increasing muscle strain amplitude (§7.1)

### 6.3. Resonance as an emergent property of the flight motor

In §6, we leveraged self-oscillation as a tool to understand motor resonance, following the classical results of Machin and Pringle [4]. The self-oscillation paradox restricts this analysis: either to low wingbeat amplitudes, or to conjectural solutions to the paradox (*e.g.*, $r \geq c$). Solving the paradox and eliminating this restriction is an avenue for future research, but even within the current framework, we are able to make one conclusion with a high degree of confidence: that *D. melanogaster* is not resonant with respect to exoskeletal elasticity (§5.2). It might be resonant, in some sense, with respect to muscular elasticity, but details are unclear. The muscular storage modulus ($E$) is elastic-like, but is optimistic to assume that it represents solely passive energy absorption [12,66]. Feedback-driven forcing functions can have a storage modulus, but store little or no energy: *e.g.*, a delayed viscous force, $F = \dot{x}(t - \delta)$, or the electronic implementation of muscular forcing devised by Lynch *et al.* [5]. In such cases, the appearance of a storage modulus is an aliasing effect [12]. Given the possibility of aliasing, the effective resonant frequencies studied in §5 may not actually be optimal: there is a need for a more pessimistic analysis of flight muscle behaviour.

As can be seen, there are several lines of analysis for future characterisations of flight motor resonance. These lines of analysis complement existing approaches that focus on alignment of experimentally-observed frequencies [34]; or other features attributed to resonance [67]. Following [5,28], we study a third aspect of flight motor resonance: resonance as an emergent property of the Dipteran flight motor, which can be characterised by modelling individual components of the flight motor (wing, exoskeleton, muscle, *etc.*); and studying the behaviour of these components as they interact. Such a characterisation involves the development both of new component-specific data-driven modelling techniques, such as the TIIR process (§3.2); and of techniques for interactive or integrative analysis, such as symbolic computation (§4). Together, these techniques can drive classical experimental results in new directions: allowing us, for instance, to extend the classical results of Machin and Pringle [4] into conditions more representative of *D. melanogaster in vivo*.

There is also scope for further extensions of this approach, *e.g.*, into forward [68], take-off [69], or manoeuvring [70] flight conditions. These conditions are likely to reflect interesting variations of the paradox: they show broadly similar levels of damping [69,70], or possibly decreased in the case of forward flight [68], but differences in inertial and dissipative loading raise the question of how the motor adapts to ensure load consistency and continued self-oscillation. Is this adaption entirely reflected in wingbeat kinematic changes, or do the structural properties of the motor adapt too? In this direction, the question of flight motor self-oscillation is connected to wider questions regarding the mechanisms of insect flight control. Remarkably, despite progress in the question of flight control [71], the decades-old question of whether the Dipteran flight motor is actually resonant still remains unresolved. However, by integrating data across the flight motor, and leveraging new modes of analysis, answers are beginning to emerge.

**Data availability statement:**

Code containing MATLAB-MuPAD symbolic computation proofs and processed results from source datasets are available on the Swedish National Data Service, at DOI: 10.5878/dq5w-x472